\documentclass[
 	screen,
 	nonacm,
]{acmart}

\usepackage{amsmath}
\usepackage{graphicx}
\usepackage{xcolor}
\usepackage{tabularx}
\usepackage{multirow}
\usepackage{hyperref}
\usepackage{subcaption}
\usepackage{enumitem}
\usepackage{fontawesome}

\DeclareMathOperator*{\dm}{dm}

\newcommand{\norm}[1]{\left\lVert#1\right\rVert}
\newcommand{\frobnorm}[1]{\norm{#1}_F}
\newcommand{\frobnormsq}[1]{\frobnorm{#1}^2}

\newcommand*{\sumContexts}{\sum_{c_1}^{l_1} \cdots \sum_{c_d}^{l_d}}

\newcommand*{\sumDims}{m + n + l}
\newcommand*{\sumDimsMD}{m + n + l_1 + \cdots + l_d}

\newcommand*{\allIndices}{u, i, c}
\newcommand*{\Xindexed}{X_{\allIndices}}
\newcommand*{\Windexed}{W_{\allIndices}}

\newcommand*{\allIndicesMD}{u, i, c_1, \ldots, c_d}
\newcommand*{\XindexedMD}{X_{\allIndicesMD}}
\newcommand*{\WindexedMD}{W_{\allIndicesMD}}

\newcommand*{\Bcontext}[1][d]{B^{\textit{(#1)}}}
\newcommand*{\Bindexed}[1]{\Bcontext[#1]_{c_#1}}

\newcommand*{\vecOne}{\vec{1}\strut}

\newcommand*{\loss}{\mathcal{L}}
\newcommand*{\lossl}[1]{\loss^{\textit{(#1)}}}

\setcopyright{rightsretained}
\copyrightyear{2023}
\acmYear{2023}

\acmConference[CARS Workshop]{Workshop on Context-Aware Recommender Systems}{September 18, 2023}{Singapore}

\begin{document}

	\title{Weighted Tensor Decompositions for Context-aware Collaborative Filtering}

	\author{Joey De Pauw}
	\email{joey.depauw@uantwerpen.be}
	\orcid{0000-0002-1417-922X}
	\affiliation{%
	  \institution{University of Antwerpen}
	  \streetaddress{Middelheimlaan 1}
	  \city{Antwerp}
	  \country{Belgium}
	}
	
	\author{Bart Goethals}
	\email{bart.goethals@uantwerpen.be}
	\orcid{0000-0001-9327-9554}
	\affiliation{%
		\institution{University of Antwerpen}
		\streetaddress{Middelheimlaan 1}
		\city{Antwerp}
		\country{Belgium}
	}
	\affiliation{%
		\institution{Monash University}
		\city{Melbourne}
		\country{Australia}
	}

	\begin{abstract}



	Over recent years it has become well accepted that user interest is not static or immutable.
	There are a variety of contextual factors, such as time of day, the weather or the user's mood, that influence the current interests of the user. Modelling approaches need to take these factors into account if they want to succeed at finding the most relevant content to recommend given the situation.
	
	A popular method for context-aware recommendation is to encode context attributes as extra dimensions of the classic user-item interaction matrix, effectively turning it into a tensor, followed by applying the appropriate tensor decomposition methods to learn missing values. However, unlike with matrix factorization, where all decompositions are essentially a product of matrices, there exist many more options for decomposing tensors by combining vector, matrix and tensor products.
	We study the most successful decomposition methods that use weighted square loss and categorize them based on their tensor structure and regularization strategy. Additionally, we further extend the pool of methods by filling in the missing combinations.

	In this paper we provide an overview of the properties of the different decomposition methods, such as their complexity, scalability, and modelling capacity. These benefits are then contrasted with the performances achieved in offline experiments to gain more insight into which method to choose depending on a specific situation and constraints.


	\end{abstract}

	\keywords{recommendation, context-aware, tensor decomposition, implicit feedback, collaborative filtering}
	
	\maketitle
	

\section{Introduction}
\label{sec:introduction}
	
	Collaborative filtering techniques are currently the most used and most studied approaches for top-k recommendation with implicit feedback. In collaborative filtering, a user's recommendations are based on the history of items they consumed and how similar they are to other users. This approach has its limitations, as often much more data is available to learn from, such as personal information of the user, metadata of the item, or contextual information about the interactions. Especially the usage of contexts is found to be indispensable to model the dynamic nature of user behaviour and to make sure the recommender system can adapt to the current needs of the user~\cite{pagano2016contextual, adomavicius2010context}.
	
	Implicit feedback data is often represented as a user-by-item binary interaction matrix where each interaction (view/click/\ldots) is represented by a \emph{one} and everything else as \emph{zero}. As the item catalogue grows, this matrix naturally becomes increasingly sparse since every user typically only observes a limited amount of items. This problem is exacerbated further as context dimensions are added,
	because each dimension exponentially increases the sparsity of the tensor.\looseness=-1
	
	Learning from sparse, positive-only data is one of the key challenges in recommendation and requires specialised learning methods such as negative sampling or importance weighting~\cite{hu2008collaborative, pan2008one} to prevent models from overfitting to the missing data~\cite{pan2008one, he2016fast}.
	Both approaches have their pros and cons: while negative sampling has more efficient training steps, convergence may be slow and highly dependent on the learning rate. Importance weighting on the other hand can use the whole data and is hence more costly, but it is often found to be more effective~\cite{chen2020efficient, he2016fast}.
	In this paper, we focus on whole-data models with weighted square loss.
%
	Weighted Matrix Factorization (WMF), also called iALS~\cite{hu2008collaborative, pan2008one}, pioneered this class of models and is still known to achieve competitive results while having highly scalable learning and prediction routines~\cite{rendle2022revisiting}. After its introduction, many extensions were proposed, among which three variants for context-aware recommender systems (CARS)~\cite{hidasi2012fast, hidasi2014factorization, chou2018fast}, where each variant uses a different tensor decomposition method.\looseness=-1

	Table~\ref{tab:models} classifies the WMF-based CARS models in terms of \emph{decomposition method}, \emph{tensor structure} and \emph{regularization strategy}. First, for tensor structure, we distinguish between models that support arbitrary dimensions and models that are limited to three dimensions. In the latter case, if there are multiple context features, they have to be stacked in one dimension. Then, the standard regularization strategy `zero' uses the same $\ell^2$-norm for all factors, pulling them all towards the origin. Our new strategy `one' regularizes the context factors towards the mathematical identity which benefits learning as per our experimental results. Finally, we consider the three most practical tensor decomposition methods: CANDECOMP/PARAFAC~(CP)~\cite{hitchcock1927expression} where each interaction is modelled by a product of vectors, Pairwise Interaction Tensor Factorization~(PITF)~\cite{rendle2010pairwise} where the sum of pairwise products of vectors is used, and Tensor Train Factorization~(TTF)~\cite{oseledets2011tensor} where user and item factors are vectors and contexts are modelled as matrices.
	The Tucker decomposition~\cite{hitchcock1927expression, tucker1966some} is not included because no WMF-based CARS algorithm with this decomposition has been published yet~\cite{frolov2017tensor}, likely because the precomputation step that leads to the efficiency of WMF, is not possible due to the use of a core tensor.\looseness=-1
	
	Six spaces in this table are already filled by previously proposed methods and we fill in four of the missing combinations that are indicated with an asterisk. Note that iTALSx and FTF are classified both under zero and one regularization for PITF as `zero' already is the mathematical identity for the factors and it hence coincides with the `one' variant.
	Also worth mentioning is the General Factorization Framework~\cite{hidasi2016general} which studies combinations of CP and PITF, and as such would fit somewhere between these rows of the table.
	Only two entries are left for multidimensional TTF, which is out of the scope of this paper and remains open for future research.
	This paper has three main contributions:
	\begin{enumerate}[topsep=0pt]
		\item Introducing Weighted Tensor Factorization (WTF), a novel WMF-based model for CARS with the TTF structure.
		\item Exploring a new regularization method for context factors that aligns more with the recommendation task.
		\item Experimental evaluation and comparison of the listed models on three publicly available datasets.
	\end{enumerate}
	
	\begin{table}
		\caption{Overview of the studied CARS models. Methods marked with asterisk (*) are our contributions.}
		\label{tab:models}
		\centering
		{
		\setlength\extrarowheight{1pt}
		\begin{tabular}{|c|cc|cc|}
			\hline
			\multicolumn{1}{|r|}{\textit{Tensor Structure:}} & \multicolumn{2}{c|}{\textbf{Up to 3D / Stacked to 3D}} & \multicolumn{2}{c|}{\textbf{Multidimensional}} \\
			\multicolumn{1}{|r|}{\textit{Regularization:}} & \textbf{zero} & \textbf{one} & \textbf{zero} & \textbf{one} \\
			\hline
			\textbf{CP}~\cite{hitchcock1927expression} & iTALSs~\cite{hidasi2012fast} & iTALSs one\textsuperscript{*} & iTALS~\cite{hidasi2012fast} & iTALS one\textsuperscript{*} \\
			\textbf{PITF}~\cite{rendle2010pairwise} & iTALSx~\cite{hidasi2014factorization} & iTALSx~\cite{hidasi2014factorization} & FTF~\cite{chou2018fast} & FTF~\cite{chou2018fast} \\
			\textbf{TTF}~\cite{oseledets2011tensor} & WTF\textsuperscript{*} & WTF one\textsuperscript{*} & -- & -- \\
			\hline
		\end{tabular}
		}
	\end{table}


\section{Models}
\label{sec:models}
	
	Let $X \in \{0, 1\}^{m \times n \times l_1 \times \cdots \times l_d}$ be the binary interaction tensor and $\hat{X} \in \mathbb{R}^{m \times n \times l_1 \times \cdots \times l_d}$ be the reconstructed values given a tensor decomposition method. Then the class of models we study all share the following loss function:
	\begin{equation}
		\loss = \frobnormsq{\sqrt{W} \odot \left(X - \hat{X} \right)} + \mathcal{R} \qquad \textrm{with } W = 1 + \alpha X
		\label{eq:generic_loss}
	\end{equation}
	Where $\odot$ denotes the elementwise product and $\frobnorm{\cdot}$ the Frobenius norm. $W$ are the weights, and $\mathcal{R}$ represents the regularization. Intuitively, this loss takes a weighted sum of the squared error for both seen and unseen interactions. However, all unseen interactions are given the same weight of one, which allows for efficient optimization~\cite{hu2008collaborative}. Positive interactions are given a constant weight of $1+\alpha$, but it can also be dependent on the strength of the signal (e.g.: rating, recency or watch duration). By using square loss, most models in this class are bi-convex and can be optimized with Alternating Least Squares (ALS)~\cite{hu2008collaborative, pan2008one}. In ALS, we learn the user factors while keeping the item and context factors constant and vice-versa for the others until convergence to a local optimum~\cite{rendle2011fast}. 
		
	

	Notice that in Equation~\eqref{eq:generic_loss}, $X$, $W$ and $\hat{X}$ are tensors of dimension $2 + d$ for multidimensional models, with $d$ the amount of context dimensions. Each context feature is added as a separate dimension, which allows dependencies between contexts to be modelled as well. In most practical cases however, there are no meaningful interactions between contexts to learn, or there is not enough data to learn them from~\cite{hidasi2012fast}. For those cases, we can reduce the model complexity by \emph{stacking} all context features into a single dimension, making it so only a three-dimensional tensor needs to be factorized.
	During training, this means that a single interaction is mapped to multiple \emph{ones} in the stacked tensor. For inference, we simply take the average of the predictions for every context. 
	
	Another important decision we made, was not to learn from missing values in our models, i.e.: there is no substitute factor that is learned for all missing values. Doing so may lead to overfitting to errors in the data that do not generalize. Instead we use a static \emph{default factor} for missing values that depends on the regularization, namely the target towards which the context factors are being pulled.\looseness=-1

	The next sections explain the loss function, update formulas (found by setting the partial derivatives to zero) and complexity for each of the three decomposition methods. We use a uniform notation with $P \in \mathbb{R}^{m \times k}$ for the user factors, $Q \in \mathbb{R}^{n \times k}$ for item factors and $\Bcontext[c]$ for contexts. Implementations in Python of the models and experiments are available:~\url{https://github.com/JoeyDP/CARS-Experiments}.
	
	\subsection{CP Decomposition: iTALS \& iTALSs}
		
		The CP decomposition models every entry as a product of vectors, leading to the following loss and update formula:
		\begin{equation*}
			 	\lossl{iTALS} = \sum_u^m \sum_i^n \sumContexts \WindexedMD \left( \XindexedMD - P_u \dm(\Bindexed{1}) \cdots \dm(\Bindexed{d})Q_i^\top \right)^2
		 		+ \lambda \left(\frobnormsq{P} + \frobnormsq{Q} + \sum_c^d\frobnormsq{\Bcontext[c] - \vecOne  \vecOne^\top} \right)
	 	\end{equation*}
		\begin{multline*}
	 		P_u^\top = \biggl( ({\Bcontext[1]}^\top \Bcontext[1]) \odot \cdots \odot ({\Bcontext[d]}^\top \Bcontext[d]) \odot (Q^\top Q) +
	 			\sum_i^n \sumContexts \WindexedMD^\prime \dm(\Bindexed{1}) \cdots \dm(\Bindexed{d})Q_i^\top Q_i \dm(\Bindexed{d}) \cdots \dm(\Bindexed{1}) + \lambda I \biggr)^{-1} \\
	 	\biggl(\sum_i^n \sumContexts \WindexedMD \XindexedMD \dm(\Bindexed{1}) \cdots \dm(\Bindexed{d})Q_i^\top \biggr)
	 	\end{multline*}	

		with $\WindexedMD = \WindexedMD^\prime + 1$. Only the update formula for the user factor is given, as the item and context factors are computed similarly. Notice that, despite our usage of \emph{diagonal matrix} notation $\dm(\cdot)$, all factors are vectors and only their element-wise product is needed in this loss, which is commutative. Hence the derivation of all factors follows the same structure, with the exception of the slightly different regularization of context factors.
		We use iTALSs to refer to the \emph{stacked} version of iTALS.
		The computational complexity for one update step of all factors is $\mathcal{O}(k^3 (\sumDimsMD) + k^2 p)$ with $p$ the amount of interactions.\looseness=-1

		\subsubsection{Regularization Towards the Identity}
		Context factors have an additional term, $- \vecOne  \vecOne^\top$, in their regularization. If we omit this part, all context factors are regularized towards zero (iTALS), and by including this term, the context factors are pulled towards the vector of ones (iTALS one). There are two reasons for taking this approach. First, the elementwise product with a vector of ones is the identity operation in this decomposition, so if a certain context factor is not informative for predicting interactions, the model is not penalized for not learning from it. Intuitively the context factor can be seen as a learned offset from the base user-item prediction. Second, we can choose the `default factor' for missing context values to be the vector of ones, without changing the magnitude of the prediction. Indeed if all known context factors are learned to be around zero, suddenly using a vector of ones in the loss will throw off the user and item factors, as the prediction will suddenly be much larger.

	
	\subsection{PITF Decomposition: iTALSx}

		In PITF all factors are also vectors. The difference is that pairwise combinations are multiplied. For three dimensions or the stacked version, we get:
		\begin{equation*}
			\lossl{iTALSx} = \sum_u^m \sum_i^n \sum_c^l \Windexed \left( \Xindexed - P_u Q_i^\top - P_u B_c^\top- Q_i B_c^\top\right)^2
			+ \lambda \left( \frobnormsq{P} + \frobnormsq{Q} + \frobnormsq{B} \right)
		\end{equation*}
		\begin{multline*}
			P_u^\top = \biggl(\sum_i^n \sum_c^l \Windexed^\prime (Q_i + B_c)^\top(Q_i + B_c) + l \cdot Q^\top Q + n \cdot B^\top B + Q^\top \vecOne \vecOne^\top B + B^\top \vecOne \vecOne^\top Q + \lambda I\biggr)^{-1} \\
			\Biggl( \sum_i^n \sum_c^l \Windexed \Xindexed (Q_i + B_c)^\top + \sum_i^n \sum_c^l \Windexed^\prime Q_i B_c^\top \cdot (Q_i + B_c)^\top + Q^\top Q B^\top \vecOne + B^\top B Q^\top \vecOne \Biggr)
		\end{multline*}	
		
		The computational complexity is equal to that of iTALS: $\mathcal{O}(k^3 (\sumDims) + k^2 p)$.
		Notice that there is no `one' variant for iTALSx, because in this decomposition, multiplication with zero vectors is actually the identity operation. I.e.\ if all context factors are zero, the prediction is solely based on the user-item part. Furthermore, we were unable to replicate the multidimensional extension of PITF for CARS, FTF~\cite{chou2018fast}, or any of its variants with reasonable effort due to missing source code and its complex model formulation. Additionally, the method scales linearly with tensor size which is not desirable.\looseness=-1
	
	\subsection{TTF Decomposition: WTF}

		Our novel contribution, Weighted Tensor Factorization~(WTF), is based on the Tensor Train Factorization~\cite{oseledets2011tensor}. Its structure is similar to iTALSs with the exception of context factors being full matrices instead of only vectors (or diagonals):\looseness=-1
		\begin{equation*}
			\lossl{WTF} = \sum_u^m \sum_i^n \sum_c^l \Windexed \left( \Xindexed - P_u \Bcontext[c] Q_i^\top \right)^2 
			+ \lambda \left(\frobnormsq{P} + \frobnormsq{Q} + \sum_c^l \frobnormsq{\Bcontext[c] - I} \right)
		\end{equation*}
		\begin{equation*}
			P_u = \left(\sum_c^l \Bcontext[c] Q^\top Q {\Bcontext[c]}^\top + \sum_i^n \sum_c^l \Windexed^\prime \Bcontext[c] Q_i^\top Q_i {\Bcontext[c]}^\top + \lambda I \right)^{-1} \biggl( \sum_i^n \sum_c^l \Windexed\Xindexed \Bcontext[c] Q_i^\top \biggr)	
		\end{equation*}

		The update equation for item factors is similar, but with $\Bcontext[c]$ transposed. For the context matrices $\Bcontext[c]$ we find:
		\begin{equation*}
			P^\top P \Bcontext[c] Q^\top Q + \sum_u^m \sum_i^n \Windexed^\prime P_u^\top P_u \Bcontext[c] Q_i^\top Q_i + \lambda \Bcontext[c] = \sum_u^m \sum_i^n \Windexed \Xindexed P_u^\top Q_i + \lambda I
		\end{equation*}
		
		This can either be solved by vectorization~\cite{lancaster1985theory} or with approximate sparse solvers, such as the conjugate gradient method~(CG)~\cite{hestenes1952methods}. The computational complexity for an update step of the user and item factors is $\mathcal{O}(k^3 (m + n) + k^2 p)$, the same as for iTALS(x). Updating all the context matrices with an exact method (vectorization) is more expensive:~$\mathcal{O}(k^6 c + k^2 p)$. However, it is found that an approximation of the update step often suffices for the algorithm to converge~\cite{takacs2011applications}. This is not surprising, as the computed value is overwritten in each iteration anyway, so computing it with high precision is not needed.
		Computing the context matrices with the CG method instead of vectorization reduces the complexity back to cubic in practice because only a small amount of CG steps (typically two or three~\cite{takacs2011applications}) are needed for adequate convergence. Note that approximate solvers can also be leveraged in the other methods to make them more efficient. For WTF however, their use is \emph{required} to make the method computationally feasible for reasonably sized datasets. There is again a `one' variant of WTF with regularization of the context factors to the identity matrix, similar to iTALS one.\looseness=-1
		
	\subsection{Frequency-based Regularization}
		To keep the notation simple, we depicted a single scalar parameter $\lambda$ for regularizing all factors. In practice, it is often found that scaling the strength of the regularization with the amount of observations improves performance~\cite{rendle2022revisiting}. Hence, we adapted the hyperparameter $\nu \in [0, 1]$ of \cite{rendle2022revisiting}, which is an exponent over the sum of weights associated with the factor in the loss. With this parameter and the global $\lambda$ scalar we compute the regularization strength per factor. For example for the user factor $P_u$ in the iTALS model the regularization becomes: $\lambda \cdot \left(\sum_i^n \sumContexts \WindexedMD \right)^\nu \norm{P_u}_2^2$.
		Here $\nu=0$ corresponds to not scaling the regularization based on frequency at all.


\section{Experiments}
\label{sec:experiments}

	\subsection{Datasets}
		The different models are compared on three publicly available datasets.
		We restrict the scope of this study to purely contextual attributes that 1) depend on the interaction and not only on the user or only on the item, and 2) are known to the system at the time of recommendation. In other words, we do not use item or user metadata for context, and preferences that cannot easily be derived, such as the time a user has available for cooking, are discarded. It should be noted that, compared to popular datasets for evaluating non-context aware recommendation, unfortunately the publicly available datasets for this task are rather small~\cite{ilarri2018datasets}. The datasets are described below and their statistics are summarized in Table~\ref{tab:datasets}.\looseness=-1
		\begin{description}[topsep=0pt]
		    \item[Frappe:] Commonly used dataset in CARS research  \cite{baltrunas2015frappe, ilarri2018datasets}. It contains app usage logs that record \emph{time of day}, \emph{weekday} and the \emph{weather}. Though it does not have many users, the context features are very descriptive.
		    \item[TripAdvisor:] A hotel rating dataset commonly used in CARS \cite{alam2016joint, ilarri2018datasets}. For context we use the \emph{state} (or country) the user is in and the \emph{trip type}.
		    \item[Food.com:] A rating and review dataset of recipes~\cite{majumder2019generating}.
		    We manually extract context features based on the date attribute, namely \emph{season} and \emph{weekday}.
		\end{description}
	
		Rating datasets were made binary by choosing ratings of 3 out of 5 and above as positive. For all datasets we only retain users with at least 3 items and for the Food.com dataset we also dropped items with less than 10 interactions. Lastly, none of the datasets contain missing values for contexts, except for Frappe where the \emph{weather} attribute has 13\% missing values.\looseness=-1
		
		\begin{table*}
			\caption{Statistics of the datasets after preprocessing. Context features indicates the amount of unique values per variable.}
			\label{tab:datasets}
			\centering
			\begin{tabular}{l|cccc}
				\toprule
				\textbf{Dataset} & \textbf{Users} & \textbf{Items} & \textbf{Interactions}  & \textbf{Context features} \\
				\midrule
				Frappe		&	816		&	4\,058	&	96\,002	&	  $7\textrm{~\faClockO} + 7\textrm{~\faCalendarO} + 7\textrm{~\faCloud}$	\\
				TripAdvisor	&	2\,362	&	2\,221	&	13\,258	&	  $79\textrm{~\faMapMarker} + 5\textrm{~\faMapO}$	\\
				Food.com 	&	22\,178	&	15\,086	&	388\,362	&	  $4\textrm{~\faSunO} + 7\textrm{~\faCalendarO}$		\\
				\bottomrule
			\end{tabular}
		
			%
			%
		\end{table*}
	
	\subsection{Experimental Setup}
		
		To keep the evaluation simple and fair, we chose the leave-one-out strategy where one interaction per user is assigned to the test set and all others to the training set. This results in disjoint training and testing interactions, but not disjoint train and test users. With this approach, our prediction step becomes easier as only one top-k list needs to be computed per user-context pair of the left-out interaction, and hence all users contribute equally to the computed metrics.
		This evaluation scenario is a good fit for simulating the use of factorization models, where user factors need to be learned beforehand. It also makes the most of the limited amount of data available.
		Furthermore, retargetting already consumed items is prevented, except for Frappe, because app usage contains repeats.
		First, hyperparameters are optimized using grid search on a single split. Then, a 5-fold cross validation on the entire dataset is performed to compute the final metrics and their standard deviation.
		
		Two standard metrics in information retrieval are used to evaluate the models: Hit Rate (HR@k) and Mean Reciprocal Rank (MRR@k). HR@k computes the ratio of true positives that are included in the top-k recommendation list (without taking rank into account). Since we only have one left-out interaction this is either one or zero per user depending on whether the left-out item was recommended. Taking the average over all users gives the percentage of users that got a relevant recommendation. MRR@k is a ranking metric that takes the rank of the first hit (true positive) in the top-k recommendation list into account by weighing its contribution with one over the rank. Its value is hence lower than or equal to the HR in our setup. As top-k list sizes we chose 5 and 20, which are typical choices for evaluating recommender systems.
%
		Alongside our CARS models, we also compare with three context-unaware baselines:
		\begin{description}[topsep=0pt]
			\item[ItemKNN~\cite{deshpande2004item}:] Item-based similarity model with cosine similarity.
			\item[EASE~\cite{steck2019embarrassingly}:] State-of-the-art autoencoder based recommender with closed-form solution.
			\item[WMF~\cite{hu2008collaborative, pan2008one, rendle2022revisiting}:] State-of-the-art matrix factorization for implicit feedback data.
		\end{description}

	\subsection{Results}
	
		\begin{table*}
			\setlength{\tabcolsep}{2pt}
			\caption{Experimental results on three datasets with best result(s) marked in bold. The left and right tables are different experiments.}
			\begin{subtable}[t]{0.56\linewidth}
				\caption{Context-unaware (Base), 3D CARS (Flat), and multidimensional CARS (MD). Standard deviation rounded to thousandths given between brackets.}
				\label{tab:results_comparative}
				\begin{tabular}{ll c cc c cc}
					\multicolumn{2}{c}{\textbf{Frappe}} && MRR@5 & MRR@20 && HR@5 & HR@20 \\ 
					\cmidrule{1-2} \cmidrule{4-5} \cmidrule{7-8}
					
					\multirow{3}{*}{Base} & ItemKNN &  & .071 (.006) & .101 (.009) &  & .155 (.007) & .451 (.020) \\
					& EASE &  & .179 (.006) & .204 (.007) &  & .345 (.011) & .584 (.014) \\
					& WMF &  & .198 (.010) & .224 (.010) &  & .368 (.010) & .619 (.011) \\
					\cmidrule{1-2} \cmidrule{4-5} \cmidrule{7-8}
					\multirow{5}{*}{Flat} & iTALSs &  & .276 (.005) & .296 (.004) &  & .456 (.006) & .638 (.008) \\
					& iTALSs one &  & \textbf{.287} (.007) & \textbf{.308} (.006) &  & .445 (.014) & .634 (.011) \\
					& iTALSx &  & .283 (.006) & .302 (.005) &  & \textbf{.467} (.009) & .639 (.007) \\
					& WTF &  & .264 (.008) & .283 (.007) &  & .416 (.015) & .591 (.016) \\
					& WTF one &  & .274 (.005) & .296 (.005) &  & .450 (.007) & \textbf{.656} (.009) \\
					\cmidrule{1-2} \cmidrule{4-5} \cmidrule{7-8}
					\multirow{2}{*}{MD}  & iTALS &  & .079 (.004) & .096 (.003) &  & .155 (.006) & .330 (.005) \\
					& iTALS one &  & .280 (.009) & .301 (.007) &  & .461 (.015) & .652 (.010) \\
					
					\bottomrule
				\end{tabular}
				\vspace{.5em}
				
				\begin{tabular}{ll c cc c cc}
					\multicolumn{2}{c}{\textbf{TripAdvisor}} && MRR@5 & MRR@20 && HR@5 & HR@20 \\ 
					\cmidrule{1-2} \cmidrule{4-5} \cmidrule{7-8}
					
					\multirow{3}{*}{Base} & ItemKNN &  & .006 (.001) & .008 (.001) &  & .011 (.001) & .040 (.002) \\
					& EASE &  & .020 (.001) & .025 (.001) &  & .037 (.002) & .085 (.004) \\
					& WMF &  & .022 (.002) & .028 (.002) &  & .040 (.004) & .102 (.002) \\
					\cmidrule{1-2} \cmidrule{4-5} \cmidrule{7-8}
					\multirow{5}{*}{Flat} & iTALSs &  & .009 (.001) & .012 (.001) &  & .018 (.002) & .056 (.003) \\
					& iTALSs one &  & .023 (.002) & .029 (.001) &  & .041 (.003) & .104 (.001) \\
					& iTALSx &  & \textbf{.026} (.002) & \textbf{.032} (.002) &  & \textbf{.047} (.004) & \textbf{.110} (.007) \\
					& WTF &  & .022 (.002) & .027 (.001) &  & .039 (.002) & .094 (.003) \\
					& WTF one &  & .025 (.002) & .031 (.002) &  & .045 (.004) & \textbf{.110} (.002) \\
					\cmidrule{1-2} \cmidrule{4-5} \cmidrule{7-8}
					\multirow{2}{*}{MD}  & iTALS &  & .006 (.002) & .009 (.002) &  & .014 (.002) & .043 (.003) \\
					& iTALS one &  & .021 (.002) & .026 (.002) &  & .039 (.004) & .096 (.002) \\
	
					\bottomrule
				\end{tabular}
				\vspace{.5em}
			
				\begin{tabular}{ll c cc c cc}
					\multicolumn{2}{c}{\textbf{Food.com}} && MRR@5 & MRR@20 && HR@5 & HR@20 \\ 
					\cmidrule{1-2} \cmidrule{4-5} \cmidrule{7-8}
					
					\multirow{3}{*}{Base} & ItemKNN &  & .010 (.001) & .012 (.001) &  & .016 (.001) & .030 (.001) \\
					& EASE &  & \textbf{.016} (.001) & .018 (.001) &  & .026 (.001) & .056 (.001) \\
					& WMF &  & .015 (.000) & \textbf{.019} (.000) &  & .026 (.001) & .063 (.001) \\
					\cmidrule{1-2} \cmidrule{4-5} \cmidrule{7-8}
					\multirow{5}{*}{Flat} & iTALSs &  & .011 (.000) & .014 (.000) &  & .021 (.001) & .053 (.001) \\
					& iTALSs one &  & .014 (.000) & .017 (.000) &  & .024 (.001) & .058 (.002) \\
					& iTALSx &  & .014 (.000) & .017 (.000) &  & .024 (.001) & .059 (.001) \\
					& WTF &  & .011 (.000) & .014 (.000) &  & .020 (.001) & .047 (.001) \\
					& WTF one &  & \textbf{.016} (.000) & \textbf{.019} (.000) &  & \textbf{.028} (.001) & \textbf{.065} (.001) \\
					\cmidrule{1-2} \cmidrule{4-5} \cmidrule{7-8}
					\multirow{2}{*}{MD}  & iTALS &  & .007 (.001) & .009 (.001) &  & .014 (.001) & .035 (.007) \\
					& iTALS one &  & \textbf{.016} (.000) & \textbf{.019} (.000) &  & \textbf{.028} (.000) & \textbf{.065} (.000) \\
					
					\bottomrule
				\end{tabular}
			\end{subtable}
			\hfill
			\vrule
			\hfill
			\begin{subtable}[t]{0.42\linewidth}
				\caption{CARS with context-unaware user-item factors. Percentage of original performance marked in brackets.}
				\label{tab:results_wmf_ca}
				\begin{tabular}{ll c ll}
					\multicolumn{1}{c}{MRR@5} & \multicolumn{1}{c}{MRR@20} && \multicolumn{1}{c}{HR@5} & \multicolumn{1}{c}{HR@20} \\ 
					\cmidrule{1-2} \cmidrule{4-5}
					\\
					\\
					
					.198 (100\%) & .224 (100\%) &  & .368 (100\%) & .619 (100\%) \\
					\cmidrule{1-2} \cmidrule{4-5}
					.198 (72\%) & .226 (76\%) &  & .362 (79\%) & .626 (98\%) \\
					.199 (69\%) & .226 (73\%) &  & .365 (82\%) & .627 (99\%) \\
					.204 (72\%) & .230 (76\%) &  & .376 (81\%) & .625 (98\%) \\
					.240 (91\%) & .265 (94\%) &  & \textbf{.417} (100\%) & \textbf{.650} (110\%) \\
					\textbf{.242} (88\%) & \textbf{.266} (90\%) &  & .416 (92\%) & .647 (99\%) \\
					\cmidrule{1-2} \cmidrule{4-5}
					.165 (209\%) & .187 (195\%) &  & .299 (193\%) & .515 (156\%) \\
					.198 (71\%) & .225 (75\%) &  & .367 (80\%) & .627 (96\%) \\
					
					\bottomrule
				\end{tabular}
				\vspace{.5em}
				
				\begin{tabular}{ll c ll}
					\multicolumn{1}{c}{MRR@5} & \multicolumn{1}{c}{MRR@20} && \multicolumn{1}{c}{HR@5} & \multicolumn{1}{c}{HR@20} \\ 
					\cmidrule{1-2} \cmidrule{4-5}
					\\
					\\
					
					.022 (100\%) & .028 (100\%) &  & .040 (100\%) & .102 (100\%) \\
					\cmidrule{1-2} \cmidrule{4-5}
					.023 (256\%) & .029 (242\%) &  & .044 (244\%) & .104 (186\%) \\
					.023 (100\%) & .028 (97\%) &  & .043 (105\%) & .105 (101\%) \\
					.024 (92\%) & .029 (91\%) &  & .044 (94\%) & \textbf{.106} (96\%) \\
					.024 (109\%) & .029 (107\%) &  & .043 (110\%) & .102 (109\%) \\
					\textbf{.025} (100\%) & \textbf{.030} (97\%) &  & \textbf{.045} (100\%) & \textbf{.106} (96\%) \\
					\cmidrule{1-2} \cmidrule{4-5}
					.023 (383\%) & .029 (322\%) &  & .044 (314\%) & .104 (242\%) \\
					.023 (110\%) & .028 (108\%) &  & .043 (110\%) & .105 (109\%) \\
					
					\bottomrule
				\end{tabular}
				\vspace{.5em}
				
				\begin{tabular}{ll c ll}
					\multicolumn{1}{c}{MRR@5} & \multicolumn{1}{c}{MRR@20} && \multicolumn{1}{c}{HR@5} & \multicolumn{1}{c}{HR@20} \\ 
					\cmidrule{1-2} \cmidrule{4-5}
					\\
					\\
					
					\textbf{.015} (100\%) & \textbf{.019} (100\%) &  & \textbf{.026} (100\%) & \textbf{.063} (100\%) \\
					\cmidrule{1-2} \cmidrule{4-5}
					.013 (118\%) & .017 (121\%) &  & .023 (110\%) & .057 (108\%) \\
					.013 (93\%) & .017 (100\%) &  & .023 (96\%) & .057 (98\%) \\
					.014 (100\%) & .017 (100\%) &  & .024 (100\%) & .058 (98\%) \\
					.014 (127\%) & .018 (129\%) &  & .025 (125\%) & .060 (128\%) \\
					.014 (88\%) & .017 (89\%) &  & .024 (86\%) & .057 (88\%) \\
					\cmidrule{1-2} \cmidrule{4-5}
					.013 (186\%) & .016 (178\%) &  & .023 (164\%) & .054 (154\%) \\
					.013 (81\%) & .017 (89\%) &  & .023 (82\%) & .057 (88\%) \\
					
					\bottomrule
				\end{tabular}
			\end{subtable}
		\end{table*}
		
		First, we compare the baselines and all CARS methods on the three datasets as shown in Table~\ref{tab:results_comparative}. Ten training steps and $k=80$ for the latent dimension are used which achieved convergence and optimal results for these relatively small datasets. We observe that Frappe shows the biggest increase in performance when using context features. For TripAdvisor a small but consistent improvement can be observed, and finally, for the Food.com dataset most models perform worse than the best baseline except for two that perform about equally well. From this, it is clear that the context of the Food.com dataset is less informative for computing recommendations. Despite these non-informative or misleading context features that appear to not generalize well, it is still interesting to see that WTF one and iTALS one can match the best baseline. Especially since the performance of all other models decreases by taking context into account. We conclude that WTF one and iTALS one are the most robust to irrelevant information. This also makes sense intuitively, as the regularization of these models pushes them towards learning offsets from the standard user-item prediction. Indeed, if there is no clear signal in the context features, their factors will not deviate from the prediction based on only the user and item.\looseness=-1
		
		Second, we find that the `one' variants of the models outperform their respective classically regularized variants on all datasets. A possible explanation for this, is that the improved regularization is more appropriate for the modelling task and in general a better match for the role contextual data plays in recommendation. Where we expect a strong influence from users to items, as both entities have unique characteristics and preferences, this is simply not the case for most context features. For example, if a user wants to use a specific app, then whether it is rainy or sunny outside or what the current time is will likely only slightly influence this preference. Modelling the context features as offsets on an already established user-item preference (rather than an equally important entity) makes more sense from this point of view.\looseness=-1
		
		An extreme case of this phenomenon can be observed when comparing iTALS with iTALS one. It is clear that iTALS is not able to model any of the three datasets, likely due to the compounding effect of treating context combinations as equally important entities. First, by modelling combinations of contexts, the data to learn from becomes even more sparse as the dimensions increase, but the amount of positive interactions remains the same. Second, because context features are modelled the same as users and items in the loss, they are given the same importance, allowing them to outweigh the user and item factors as the number of context dimensions grows.
		Except for specific situations such as user cold-start, this is likely not desirable for the recommendation task.
		iTALS one prevents this behaviour by encouraging the context factors to only model offsets to the user and item predictions.
				
		For the Frappe dataset specifically, the poor performance of iTALS can also in part be explained by the missing context features. Since factors of missing features are set to the zero vector, the entire score for all items will be zero and no sensible ranking can be made. iTALS one and iTALSs do not have this problem because the missing context factor is set to one in the former and the average score is taken with other contexts in the latter.
		
		Third, to investigate flattening context features over modelling them multidimensionally we can compare iTALSs one with iTALS one. On the Frappe and TripAdvisor datasets, we find no improvement to modelling combinations of contexts. Again, this can mainly be attributed to the data becoming exponentially more sparse as the number of dimensions grows. In most cases it also seems more intuitive to model contexts as independent, because we do not expect the information to drastically change when contexts occur together. For example, it may be raining in the evening or raining in the morning but using the offset of raining plus the offset of the time will likely be a good proxy for both cases. If dependencies were present in the data (and if there would be enough data to learn them from), then we can expect multi-dimensional models to outperform their flattened variants.
		
		Finally, the iTALSx model achieves good results across all datasets.
		Only on the Food.com dataset we can see it is outperformed by WTF one and iTALS one, indicating these models may have a higher modelling capacity or could be less prone to noise. More data is needed to support these hypotheses.
		


	\subsection{Results with Context-unaware User-item Factors}
		
		In a second experiment, we investigate whether context factors can be learned post-hoc with pre-learned user and item factors. Given that many recommender systems already use embeddings, if the context factors can be learned after the training procedure that is already in place, they are easier to integrate into the system. Additionally, the training cost is reduced as the context factors only need one (or very few for multi-dimensional) iteration(s) and the learning of user and item factors does not need to take context into account.
			
		The results of this experiment are reported in Table~\ref{tab:results_wmf_ca} where we used the optimal factors of WMF for each dataset followed by only learning the context factors of each method in one iteration (three for iTALS and iTALS one to learn dependencies between contexts). Hyperparameters were optimized again for learning the contexts. Though this training procedure is significantly faster, it is clear from the results that not all context information can be extracted in this setup. The methods that perform better than when they were trained with context from the start are the ones that got results worse than the WMF baseline and were now able to achieve similar or slightly higher metrics than the context-unaware baseline. However none of the methods can match the optimal results of the previous experiment as can very clearly be seen for the Frappe dataset and also for TripAdvisor to a lesser extent. For the Food.com dataset there is even less information to be learned from contexts and all results are close to the WMF baseline.
	
		If we look at which method is best suited for post-hoc learning, we find that WTF and WTF one are able to make the best of this situation. This is not surprising, given the fact their context factors contain more weights and are able to model more complex `transformations' between the user and item factors. Nevertheless, there remains a benefit to learning user and item factors with contexts in the loss because the context under which a user consumed an item can influence those factors in a significant way. For example, if two users consumed the same items but under different contexts, their factors will converge to the same optimum, and even the best and most descriptive context factor cannot separate them post-hoc.\looseness=-1


\section{Conclusions}
\label{sec:conclusions}
	
	Tensor decompositions prove to be a powerful and intuitive tool for context-aware recommendation. Their linear scaling with the number of users, items and contexts, and easy parallellization allow them to remain computationally efficient while learning from the entire dataset without need for negative-sampling.
	In this paper, first we introduce a new regularization strategy that is more appropriate for the recommendation task. Second, we derive a new model based on the TTF decomposition. Third, we compare different decomposition methods, tensor structures and regularization strategies for CARS and demonstrate the effect those three choices have depending on the use case.
	If many context features are available that potentially depend on each other, then a multidimensional model with `one' regularization likely works best.
	Otherwise, if there are few context variables that are very rich, a model like `WTF one', that learns powerful context factors, is more applicable. Understanding the context and requirements of the recommender system is key to supplying good recommendations, as there is no single method that excels at every use case. This paper serves as a practical guide for the trade-offs to consider with weighted tensor decompositions for CARS.

	\begin{acks}
		This work was supported by the \grantsponsor{fwo}{Research Foundation --- Flanders (FWO)}{https://www.fwo.be/} [\grantnum{fwo}{11E5921N} to J. De Pauw].
	\end{acks}

	
	\bibliographystyle{ACM-Reference-Format}
	\bibliography{references}


\begin{thebibliography}{26}


\ifx \showCODEN    \undefined \def \showCODEN     #1{\unskip}     \fi
\ifx \showDOI      \undefined \def \showDOI       #1{#1}\fi
\ifx \showISBNx    \undefined \def \showISBNx     #1{\unskip}     \fi
\ifx \showISBNxiii \undefined \def \showISBNxiii  #1{\unskip}     \fi
\ifx \showISSN     \undefined \def \showISSN      #1{\unskip}     \fi
\ifx \showLCCN     \undefined \def \showLCCN      #1{\unskip}     \fi
\ifx \shownote     \undefined \def \shownote      #1{#1}          \fi
\ifx \showarticletitle \undefined \def \showarticletitle #1{#1}   \fi
\ifx \showURL      \undefined \def \showURL       {\relax}        \fi
\providecommand\bibfield[2]{#2}
\providecommand\bibinfo[2]{#2}
\providecommand\natexlab[1]{#1}
\providecommand\showeprint[2][]{arXiv:#2}

\bibitem[Adomavicius and Tuzhilin(2010)]%
        {adomavicius2010context}
\bibfield{author}{\bibinfo{person}{Gediminas Adomavicius} {and}
  \bibinfo{person}{Alexander Tuzhilin}.} \bibinfo{year}{2010}\natexlab{}.
\newblock \showarticletitle{Context-aware recommender systems}.
\newblock In \bibinfo{booktitle}{\emph{Recommender systems handbook}}.
  \bibinfo{publisher}{Springer}, \bibinfo{pages}{217--253}.
\newblock


\bibitem[Alam et~al\mbox{.}(2016)]%
        {alam2016joint}
\bibfield{author}{\bibinfo{person}{Md~Hijbul Alam}, \bibinfo{person}{Woo-Jong
  Ryu}, {and} \bibinfo{person}{SangKeun Lee}.} \bibinfo{year}{2016}\natexlab{}.
\newblock \showarticletitle{Joint multi-grain topic sentiment: modeling
  semantic aspects for online reviews}.
\newblock \bibinfo{journal}{\emph{Information Sciences}}  \bibinfo{volume}{339}
  (\bibinfo{year}{2016}), \bibinfo{pages}{206--223}.
\newblock


\bibitem[Baltrunas et~al\mbox{.}(2015)]%
        {baltrunas2015frappe}
\bibfield{author}{\bibinfo{person}{Linas Baltrunas}, \bibinfo{person}{Karen
  Church}, \bibinfo{person}{Alexandros Karatzoglou}, {and}
  \bibinfo{person}{Nuria Oliver}.} \bibinfo{year}{2015}\natexlab{}.
\newblock \showarticletitle{Frappe: Understanding the usage and perception of
  mobile app recommendations in-the-wild}.
\newblock \bibinfo{journal}{\emph{arXiv preprint arXiv:1505.03014}}
  (\bibinfo{year}{2015}).
\newblock


\bibitem[Chen et~al\mbox{.}(2020)]%
        {chen2020efficient}
\bibfield{author}{\bibinfo{person}{Chong Chen}, \bibinfo{person}{Min Zhang},
  \bibinfo{person}{Yongfeng Zhang}, \bibinfo{person}{Yiqun Liu}, {and}
  \bibinfo{person}{Shaoping Ma}.} \bibinfo{year}{2020}\natexlab{}.
\newblock \showarticletitle{Efficient neural matrix factorization without
  sampling for recommendation}.
\newblock \bibinfo{journal}{\emph{ACM Transactions on Information Systems
  (TOIS)}} \bibinfo{volume}{38}, \bibinfo{number}{2} (\bibinfo{year}{2020}),
  \bibinfo{pages}{1--28}.
\newblock


\bibitem[Chou et~al\mbox{.}(2018)]%
        {chou2018fast}
\bibfield{author}{\bibinfo{person}{Szu-Yu Chou},
  \bibinfo{person}{Jyh-Shing~Roger Jang}, {and} \bibinfo{person}{Yi-Hsuan
  Yang}.} \bibinfo{year}{2018}\natexlab{}.
\newblock \showarticletitle{Fast tensor factorization for large-scale
  context-aware recommendation from implicit feedback}.
\newblock \bibinfo{journal}{\emph{IEEE Transactions on Big Data}}
  \bibinfo{volume}{6}, \bibinfo{number}{1} (\bibinfo{year}{2018}),
  \bibinfo{pages}{201--208}.
\newblock


\bibitem[Deshpande and Karypis(2004)]%
        {deshpande2004item}
\bibfield{author}{\bibinfo{person}{Mukund Deshpande} {and}
  \bibinfo{person}{George Karypis}.} \bibinfo{year}{2004}\natexlab{}.
\newblock \showarticletitle{Item-Based Top-N Recommendation Algorithms}.
\newblock \bibinfo{journal}{\emph{ACM Trans. Inf. Syst.}} \bibinfo{volume}{22},
  \bibinfo{number}{1} (\bibinfo{date}{Jan.} \bibinfo{year}{2004}),
  \bibinfo{pages}{143–177}.
\newblock
\showISSN{1046-8188}
\urldef\tempurl%
\url{https://doi.org/10.1145/963770.963776}
\showDOI{\tempurl}


\bibitem[Frolov and Oseledets(2017)]%
        {frolov2017tensor}
\bibfield{author}{\bibinfo{person}{Evgeny Frolov} {and} \bibinfo{person}{Ivan
  Oseledets}.} \bibinfo{year}{2017}\natexlab{}.
\newblock \showarticletitle{Tensor methods and recommender systems}.
\newblock \bibinfo{journal}{\emph{Wiley Interdisciplinary Reviews: Data Mining
  and Knowledge Discovery}} \bibinfo{volume}{7}, \bibinfo{number}{3}
  (\bibinfo{year}{2017}), \bibinfo{pages}{e1201}.
\newblock


\bibitem[He et~al\mbox{.}(2016)]%
        {he2016fast}
\bibfield{author}{\bibinfo{person}{Xiangnan He}, \bibinfo{person}{Hanwang
  Zhang}, \bibinfo{person}{Min-Yen Kan}, {and} \bibinfo{person}{Tat-Seng
  Chua}.} \bibinfo{year}{2016}\natexlab{}.
\newblock \showarticletitle{Fast matrix factorization for online recommendation
  with implicit feedback}. In \bibinfo{booktitle}{\emph{Proceedings of the 39th
  International ACM SIGIR conference on Research and Development in Information
  Retrieval}}. \bibinfo{pages}{549--558}.
\newblock


\bibitem[Hestenes and Stiefel(1952)]%
        {hestenes1952methods}
\bibfield{author}{\bibinfo{person}{Magnus~R Hestenes} {and}
  \bibinfo{person}{Eduard Stiefel}.} \bibinfo{year}{1952}\natexlab{}.
\newblock \showarticletitle{Methods of conjugate gradients for solving}.
\newblock \bibinfo{journal}{\emph{Journal of research of the National Bureau of
  Standards}} \bibinfo{volume}{49}, \bibinfo{number}{6} (\bibinfo{year}{1952}),
  \bibinfo{pages}{409}.
\newblock


\bibitem[Hidasi(2014)]%
        {hidasi2014factorization}
\bibfield{author}{\bibinfo{person}{Bal{\'a}zs Hidasi}.}
  \bibinfo{year}{2014}\natexlab{}.
\newblock \showarticletitle{Factorization models for context-aware
  recommendations}.
\newblock \bibinfo{journal}{\emph{Infocommun J VI (4)}} (\bibinfo{year}{2014}),
  \bibinfo{pages}{27--34}.
\newblock


\bibitem[Hidasi and Tikk(2012)]%
        {hidasi2012fast}
\bibfield{author}{\bibinfo{person}{Bal{\'a}zs Hidasi} {and}
  \bibinfo{person}{Domonkos Tikk}.} \bibinfo{year}{2012}\natexlab{}.
\newblock \showarticletitle{Fast ALS-based tensor factorization for
  context-aware recommendation from implicit feedback}. In
  \bibinfo{booktitle}{\emph{Joint European Conference on Machine Learning and
  Knowledge Discovery in Databases}}. Springer, \bibinfo{pages}{67--82}.
\newblock


\bibitem[Hidasi and Tikk(2016)]%
        {hidasi2016general}
\bibfield{author}{\bibinfo{person}{Bal{\'a}zs Hidasi} {and}
  \bibinfo{person}{Domonkos Tikk}.} \bibinfo{year}{2016}\natexlab{}.
\newblock \showarticletitle{General factorization framework for context-aware
  recommendations}.
\newblock \bibinfo{journal}{\emph{Data Mining and Knowledge Discovery}}
  \bibinfo{volume}{30}, \bibinfo{number}{2} (\bibinfo{year}{2016}),
  \bibinfo{pages}{342--371}.
\newblock


\bibitem[Hitchcock(1927)]%
        {hitchcock1927expression}
\bibfield{author}{\bibinfo{person}{Frank~L Hitchcock}.}
  \bibinfo{year}{1927}\natexlab{}.
\newblock \showarticletitle{The expression of a tensor or a polyadic as a sum
  of products}.
\newblock \bibinfo{journal}{\emph{Journal of Mathematics and Physics}}
  \bibinfo{volume}{6}, \bibinfo{number}{1-4} (\bibinfo{year}{1927}),
  \bibinfo{pages}{164--189}.
\newblock


\bibitem[Hu et~al\mbox{.}(2008)]%
        {hu2008collaborative}
\bibfield{author}{\bibinfo{person}{Yifan Hu}, \bibinfo{person}{Yehuda Koren},
  {and} \bibinfo{person}{Chris Volinsky}.} \bibinfo{year}{2008}\natexlab{}.
\newblock \showarticletitle{Collaborative filtering for implicit feedback
  datasets}. In \bibinfo{booktitle}{\emph{2008 Eighth IEEE International
  Conference on Data Mining}}. Ieee, \bibinfo{pages}{263--272}.
\newblock


\bibitem[Ilarri et~al\mbox{.}(2018)]%
        {ilarri2018datasets}
\bibfield{author}{\bibinfo{person}{Sergio Ilarri}, \bibinfo{person}{Raquel
  Trillo-Lado}, {and} \bibinfo{person}{Ram{\'o}n Hermoso}.}
  \bibinfo{year}{2018}\natexlab{}.
\newblock \showarticletitle{Datasets for context-aware recommender systems:
  Current context and possible directions}. In \bibinfo{booktitle}{\emph{2018
  IEEE 34th International Conference on Data Engineering Workshops (ICDEW)}}.
  IEEE, \bibinfo{pages}{25--28}.
\newblock


\bibitem[Lancaster and Tismenetsky(1985)]%
        {lancaster1985theory}
\bibfield{author}{\bibinfo{person}{Peter Lancaster} {and}
  \bibinfo{person}{Miron Tismenetsky}.} \bibinfo{year}{1985}\natexlab{}.
\newblock \bibinfo{booktitle}{\emph{The theory of matrices: with
  applications}}.
\newblock \bibinfo{publisher}{Elsevier}.
\newblock


\bibitem[Majumder et~al\mbox{.}(2019)]%
        {majumder2019generating}
\bibfield{author}{\bibinfo{person}{Bodhisattwa~Prasad Majumder},
  \bibinfo{person}{Shuyang Li}, \bibinfo{person}{Jianmo Ni}, {and}
  \bibinfo{person}{Julian McAuley}.} \bibinfo{year}{2019}\natexlab{}.
\newblock \showarticletitle{Generating personalized recipes from historical
  user preferences}.
\newblock \bibinfo{journal}{\emph{arXiv preprint arXiv:1909.00105}}
  (\bibinfo{year}{2019}).
\newblock


\bibitem[Oseledets(2011)]%
        {oseledets2011tensor}
\bibfield{author}{\bibinfo{person}{Ivan~V Oseledets}.}
  \bibinfo{year}{2011}\natexlab{}.
\newblock \showarticletitle{Tensor-train decomposition}.
\newblock \bibinfo{journal}{\emph{SIAM Journal on Scientific Computing}}
  \bibinfo{volume}{33}, \bibinfo{number}{5} (\bibinfo{year}{2011}),
  \bibinfo{pages}{2295--2317}.
\newblock


\bibitem[Pagano et~al\mbox{.}(2016)]%
        {pagano2016contextual}
\bibfield{author}{\bibinfo{person}{Roberto Pagano}, \bibinfo{person}{Paolo
  Cremonesi}, \bibinfo{person}{Martha Larson}, \bibinfo{person}{Bal{\'a}zs
  Hidasi}, \bibinfo{person}{Domonkos Tikk}, \bibinfo{person}{Alexandros
  Karatzoglou}, {and} \bibinfo{person}{Massimo Quadrana}.}
  \bibinfo{year}{2016}\natexlab{}.
\newblock \showarticletitle{The contextual turn: From context-aware to
  context-driven recommender systems}. In \bibinfo{booktitle}{\emph{Proceedings
  of the 10th ACM conference on recommender systems}}.
  \bibinfo{pages}{249--252}.
\newblock


\bibitem[Pan et~al\mbox{.}(2008)]%
        {pan2008one}
\bibfield{author}{\bibinfo{person}{Rong Pan}, \bibinfo{person}{Yunhong Zhou},
  \bibinfo{person}{Bin Cao}, \bibinfo{person}{Nathan~N Liu},
  \bibinfo{person}{Rajan Lukose}, \bibinfo{person}{Martin Scholz}, {and}
  \bibinfo{person}{Qiang Yang}.} \bibinfo{year}{2008}\natexlab{}.
\newblock \showarticletitle{One-class collaborative filtering}. In
  \bibinfo{booktitle}{\emph{2008 Eighth IEEE International Conference on Data
  Mining}}. IEEE, \bibinfo{pages}{502--511}.
\newblock


\bibitem[Rendle et~al\mbox{.}(2011)]%
        {rendle2011fast}
\bibfield{author}{\bibinfo{person}{Steffen Rendle}, \bibinfo{person}{Zeno
  Gantner}, \bibinfo{person}{Christoph Freudenthaler}, {and}
  \bibinfo{person}{Lars Schmidt-Thieme}.} \bibinfo{year}{2011}\natexlab{}.
\newblock \showarticletitle{Fast context-aware recommendations with
  factorization machines}. In \bibinfo{booktitle}{\emph{Proceedings of the 34th
  international ACM SIGIR conference on Research and development in Information
  Retrieval}}. \bibinfo{pages}{635--644}.
\newblock


\bibitem[Rendle et~al\mbox{.}(2022)]%
        {rendle2022revisiting}
\bibfield{author}{\bibinfo{person}{Steffen Rendle}, \bibinfo{person}{Walid
  Krichene}, \bibinfo{person}{Li Zhang}, {and} \bibinfo{person}{Yehuda Koren}.}
  \bibinfo{year}{2022}\natexlab{}.
\newblock \showarticletitle{Revisiting the performance of ials on item
  recommendation benchmarks}. In \bibinfo{booktitle}{\emph{Proceedings of the
  16th ACM Conference on Recommender Systems}}. \bibinfo{pages}{427--435}.
\newblock


\bibitem[Rendle and Schmidt-Thieme(2010)]%
        {rendle2010pairwise}
\bibfield{author}{\bibinfo{person}{Steffen Rendle} {and} \bibinfo{person}{Lars
  Schmidt-Thieme}.} \bibinfo{year}{2010}\natexlab{}.
\newblock \showarticletitle{Pairwise interaction tensor factorization for
  personalized tag recommendation}. In \bibinfo{booktitle}{\emph{Proceedings of
  the third ACM international conference on Web search and data mining}}.
  \bibinfo{pages}{81--90}.
\newblock


\bibitem[Steck(2019)]%
        {steck2019embarrassingly}
\bibfield{author}{\bibinfo{person}{Harald Steck}.}
  \bibinfo{year}{2019}\natexlab{}.
\newblock \showarticletitle{Embarrassingly shallow autoencoders for sparse
  data}. In \bibinfo{booktitle}{\emph{The World Wide Web Conference}}.
  \bibinfo{pages}{3251--3257}.
\newblock


\bibitem[Tak{\'a}cs et~al\mbox{.}(2011)]%
        {takacs2011applications}
\bibfield{author}{\bibinfo{person}{G{\'a}bor Tak{\'a}cs},
  \bibinfo{person}{Istv{\'a}n Pil{\'a}szy}, {and} \bibinfo{person}{Domonkos
  Tikk}.} \bibinfo{year}{2011}\natexlab{}.
\newblock \showarticletitle{Applications of the conjugate gradient method for
  implicit feedback collaborative filtering}. In
  \bibinfo{booktitle}{\emph{Proceedings of the fifth ACM conference on
  Recommender systems}}. \bibinfo{pages}{297--300}.
\newblock


\bibitem[Tucker(1966)]%
        {tucker1966some}
\bibfield{author}{\bibinfo{person}{Ledyard~R Tucker}.}
  \bibinfo{year}{1966}\natexlab{}.
\newblock \showarticletitle{Some mathematical notes on three-mode factor
  analysis}.
\newblock \bibinfo{journal}{\emph{Psychometrika}} \bibinfo{volume}{31},
  \bibinfo{number}{3} (\bibinfo{year}{1966}), \bibinfo{pages}{279--311}.
\newblock


\end{thebibliography}

\end{document}